\documentclass[aps,prl,twocolumn,showpacs,superscriptaddress]{revtex4-1}

\usepackage{amsmath}
\usepackage{amssymb}
\usepackage{bm}
\usepackage{braket}
\usepackage[dvipdfmx]{graphicx}
\usepackage{hyperref}
\usepackage{tikz}
\usepackage{url}

\hypersetup{
    colorlinks=true,
    linkcolor=blue,
    filecolor=magenta,      
    urlcolor=cyan,
    }
    
\allowdisplaybreaks[1]

\begin{document}

\title{Exact solution of  the macroscopic fluctuation theory for the symmetric exclusion process}
\author{Kirone Mallick}
\email{kirone.mallick@ipht.fr}
\affiliation{%
Institut de Physique Théorique, CEA, CNRS, Université Paris–Saclay, F–91191 Gif-sur-Yvette cedex, France
}
%\footnote{Institut de Physique Théorique, CEA, CNRS, Université Paris–Saclay, F–91191 Gif-sur-Yvette cedex, France,  E-mail: kirone.mallick@ipht.fr}
\author{
Hiroki Moriya}
\email{hmoriya@stat.phys.titech.ac.jp}
%\footnote{Department of physics, Tokyo Institute of Technology, E-mail: hmoriya@stat.phys.titech.ac.jp} 
\author{
Tomohiro Sasamoto} %\footnote{Department of physics, Tokyo Institute of Technology, E-mail: sasamoto@phys.titech.ac.jp}}
\email{sasamoto@phys.titech.ac.jp}
\affiliation{%
Department of physics, Tokyo Institute of Technology, Tokyo 152-8551, Japan
}%
\date{\today}

\begin{abstract}
We present the first exact solution for  the time dependent  equations  of the macroscopic fluctuation theory (MFT) for the  symmetric simple exclusion process by combining a generalization of the canonical Cole-Hopf  transformation  with  the inverse scattering method. 
For the step initial condition with two densities, the associated Riemann-Hilbert problem is solved to determine exactly the optimal density profile and the response field which produce a required fluctuation, both at initial and final times. The large deviation function of the current is  derived and  coincides  with the formula obtained previously by microscopic calculations. This  provides the first analytic confirmation of the validity of the  MFT for an interacting model  in the time dependent regime. 
\end{abstract}

\maketitle

A fundamental difference between equilibrium and non-equilibrium physics is that a general law-- the Boltzmann-Gibbs canonical distribution --  exists in the former case. Moreover, dynamical fluctuations in the vicinity of equilibrium and linear response theory are well understood thanks to the Onsager-Machlup functional \cite{OnsMach1,OnsMach2}. However, it is now  widely believed that large deviation functions could  play an overarching role for systems far from  equilibrium 
\cite{Varadhan1984,TOUCHETTE,LebowitzSpohn1999} and 
their study has  become a major focus
of contemporary statistical mechanics  \cite{Derrida2007,Derrida2011,TouchetteHARRIS,GianniBrazil}. In a series of seminal works starting from the early 2000's, G. Jona-Lasinio and his collaborators proposed a non-linear action functional that encodes the fluctuations and the large deviations for a wide class of diffusive systems out of equilibrium. This theory, known as the Macroscopic Fluctuation Theory (MFT) \cite{Bertini2001,Bertini2002,Bodineau2004,Bodineau05,Bertini2015,Jona-Lasinio2014} posits a variational principle that determines, in a   diffusive system, the dominant optimal evolution to produce  a required fluctuation. In essence, the problem amounts to solving a set of two coupled non-linear partial differential equations (PDE's) with mixed, non-local, initial and final conditions: the MFT equations.

Another field  of research in non-equilibrium physics is to analyze microscopic interacting particle processes  that display hydrodynamic behavior on the macroscopic scale \cite{KLS1984,Spohn1991,SchutzReview}.
One of the simplest and  fundamental models is the symmetric exclusion process (SEP), in which particles on a lattice perform symmetric random walks subject to hard-core exclusion. Together with its driven version, the exclusion process plays the role of a paradigm in many domains, as 
the `simplest non-equilibrium model'
\cite{DerridaPhysREP,CHOU2011}. Many
results about exclusion processes  have been obtained analytically,
leading to  significant information about general properties of  non-equilibrium systems \cite{Derrida2007}. From the very beginning, the SEP has been a major benchmark to build and investigate the MFT \cite{Derrida-Leb01,Derrida-Leb02a,Derrida-Leb02b,Bertini2003SEP,Gerschenfeld2009Bethe,Gerschenfeld2009,IMSprl,IMScmp}.
We note that, in probability theory, a large deviation principle for SEP
had already been established  in 1989 by Kipnis, Olla, and Varadhan with a variational principle related to MFT
\cite{Kipnis89} (See also  \cite{Spohn1991,SethuramanVaradhan2013}). 

Exact  results for large deviation properties of SEP in the non-stationary
regime are quite limited. The  large deviation of the total current  through  the origin was derived  by  Bethe ansatz \cite{Gerschenfeld2009Bethe} and the full distribution of  
a tagged  particle position  was obtained in 
\cite{IMSprl,IMScmp},  using techniques from integrable probabilities 
\cite{Schutz1997,TW2008a,BCS2014,BC2014}.  However, the  extension of these approaches to time-dependent observables, such as the optimal fluctuation history of the process, appears to be out of reach. This information could be extracted from the time-dependent solutions of the MFT equations if only one could solve them: this seems to be a formidable task, since  
only stationary or perturbative solutions of the MFT were  found for SEP \cite{Bertini2003SEP,Bodineau07,Gerschenfeld2009,Krapivsky2012,KMS2014,Meerson13,Sasorov2014}.
 
Yet, it has been suspected for some stochastic processes that optimal path equations could be 'classically' integrable: this was explicitly recognized by the authors of \cite{Kamenev16b} for the Kardar-Parisi-Zhang (KPZ) equation with weak noise, a problem  solved by the Inverse Scattering Method (ISM) in 2021 \cite{LeDoussal2021a,LeDoussal2021b}. More recently, the full statistics of non-stationary heat transfer in the Kipnis-Machioro-Presutti (KMP) model has been calculated in \cite{Bettelheim2021} using again the ISM: this must be hailed as the first analytical solution of the MFT equations for a specific (and not microscopically integrable) model, with very special boundary conditions. 

In the meanwhile, Grabsch et al. \cite{Benichou21c} made a major breakthrough in the understanding of large deviations in single file systems such as the SEP. Without using neither integrability nor the MFT, they intuited and unveiled recursively a closed equation for the final optimal profile, allowing them to determine that profile and the corresponding large deviations
(see  \cite{Illien2013,Benichou21a,Benichou21b} for precursory works in the same group).
 
We  present here  a general scheme to resolve analytically the MFT equations for SEP on the infinite line. We  devise a non-local transformation that maps  these  MFT equations to the classically integrable Ablowitz-Kaup-Newell-Segur (AKNS) system, that we analyze by ISM.
For the step initial condition with two densities, 
the calculation of the scattering data leads to a solvable
Riemann-Hilbert problem,  
allowing us to determine  analytically
the density profiles and the response  fields both 
at initial and final times.
By retrieving the cumulant generating function of the current, the relevance of the MFT equations in the time dependent regime is confirmed.

\begin{figure}[ht]
    \centering
        \includegraphics[scale=0.5]{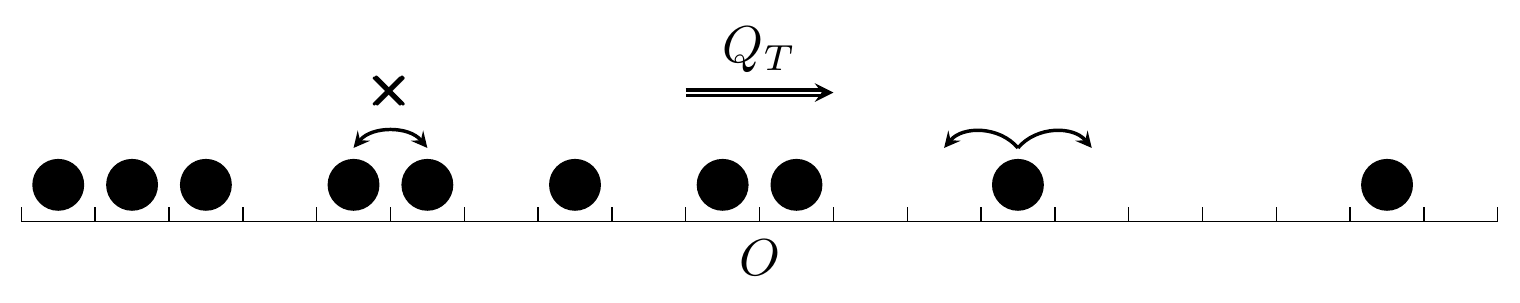}
    \caption{The symmetric simple exclusion process}
    \label{fig:SEP}
\end{figure}

The SEP is a continuous time interacting particles Markov process in which each particle is located on a discrete site labeled by an integer $x \in {\mathbb Z}$ and can hop to its right or left nearest neighboring site with unit rate. Due to the volume exclusion, jumps to an occupied site are forbidden (See Fig. \ref{fig:SEP}). We consider the time-integrated current $Q_T$,  given by
the total number of particles that have  jumped from $0$ to $1$ minus the total number of particles that have jumped from $1$ to $0$ during the time interval $(0,T)$.
In the long time limit, the current $Q_T$ satisfies a large-deviation principle
\begin{equation}
 {\rm Prob}\left( \frac{Q_T}{\sqrt{T}} = q  \right)
  \simeq  \exp[-\sqrt{T} \Phi(q)] \,  
\label{def1:PHI}
\end{equation}
with a  large deviation function $\Phi(q)$ (the $\sqrt{T}$ scaling 
   is a footprint of anomalous diffusion in single-file systems
   \cite{krapivskyBook}). The cumulant generating  function $\mu(\lambda)$
   of the current  $Q_T$ is defined as 
\begin{equation}
   \langle e^{\lambda Q_T} \rangle \simeq  e^{\sqrt{T} \mu(\lambda)} 
 \quad \hbox{ for}  ~T \to \infty \, 
\label{def:mu}
\end{equation}
where $\lambda$ is a real parameter (or  fugacity)  conditioning the  total
current  $q$ during the time interval $(0,T)$.
The functions $\Phi(q)$ and  $\mu(\lambda)$ are  Legendre transforms of each other. 
As already mentioned %in introduction
above, $\mu$ was calculated explicitly for the step initial condition in
\cite{Gerschenfeld2009Bethe}.

The MFT describes the evolution of the system in terms of two coupled fields defined on a mesoscopic scale: the
density  $\rho(x,t)$  and the auxiliary response field $H(x,t)$
(that can be interpreted as a dynamically generated local drift).
A dynamical action  is ascribed
to each history of the system.
In the long time limit, the   extremal action principle
determines the optimal path history that  produces  a required fluctuation
and  expresses  its probability  at the level of large deviations.
In the hydrodynamic limit,  the time-integrated current $Q_T$ is  given by
\begin{equation}
  Q_T=\int_0^\infty[\rho(x,T)-\rho(x,0)] dx, 
  \label{Qcontinuous}
\end{equation}
and $\mu(\lambda) \sqrt{T} $ 
is given by the maximum of the  functional 
\begin{equation}
  S[\rho,H] =  \lambda Q_T  -\mathcal{F}_0[\rho(x,0)]
  - \int_{0}^{T} dt \int_{-\infty}^\infty dx (H\partial_t \rho-
  \mathcal{H}),
\label{Action}
\end{equation}
where 
$ \mathcal{H}[\rho,H]=\frac12\sigma(\rho)(\partial_xH)^2-
(\partial_x\rho)(\partial_xH)$ with
$\sigma(\rho)=2\rho(1-\rho)$
is the MFT-Hamiltonian.
The  initial free energy is  given by 
\begin{equation}
  \mathcal{F}_0[\rho(x,0)] =  \int_{-\infty}^\infty dx
  \int_{\bar\rho(x)}^{\rho(x,0)} dr  \frac{\rho(x,0) - r}{r(1 -r)},
\end{equation}
for a Bernoulli initial state with density $\bar{\rho}(x)$. 

The dominant path  maximizing  the  action (\ref{Action})
satisfies the MFT equations that couple two
non-random optimal fields, the 
density $\rho(x,t)$ and the response  $H(x,t)$:
\begin{align}
\partial_t\rho &= \partial_x[\partial_x\rho- \sigma(\rho)\partial_xH], 
\label{eq:MFTrho}\\
\partial_tH &= -\partial_x^2H-\frac{\sigma'(\rho)}{2}(\partial_xH)^2.
\label{eq:MFTH}
\end{align}
These  equations must be solved with the following conditions at the initial and the final times:
\begin{align}
H(x,T) &= \lambda\theta(x), 
\label{eq:FCforH}\\
H(x,0) &= \lambda\theta(x)+f'(\rho(x,0))-f'(\bar\rho(x)),
\label{eq:ICforH}
\end{align}
where $f'(\rho) = \log\frac{\rho}{1-\rho}$ is the derivative of the free energy with respect to the density \cite{Gerschenfeld2009,KMS_tagged}. The MFT equations are a set of non-linear coupled PDEs that evolve in opposite time directions, with a major complication  due to the two-time mixed boundary conditions, rendering  numerical simulations arduous \cite{Vilenkin14,Sasorov2014}.
In this work, we show that the MFT equations,  
(\ref{eq:MFTrho}, \ref{eq:MFTH}) with conditions (\ref{eq:FCforH}) and (\ref{eq:ICforH}), are
integrable in the classical sense  after a nonlinear transformation and solve them analytically.
This framework can be stated for general initial condition but we focus here on 
the two-sided Bernoulli initial condition: at $t=0$ all sites are independent, a site with negative label is occupied with probability $\rho_-$ and a site on the positive side is occupied with probability $\rho_+$.
As the initial condition fluctuates, this set-up is said to be annealed
and the mean density profile at initial time is given by $\bar\rho(x)=\rho_-\theta(-x)+\rho_+\theta(x)$ (where $\theta(x)$ is the Heaviside theta function).

The ISM \cite{Ablowitz1981,faddeev1987}, often presented as a non-linear analog of the Fourier Transform, lies at the heart of classical integrability and has led to exact solutions of PDEs in various domains of physics, such as the Korteweg-de Vries equation or the nonlinear Schr\"odinger equation.
As already mentioned, the ISM has been used in recent works to solve the short time KPZ equation \cite{LeDoussal2021a,LeDoussal2021b} and the KMP model \cite{Bettelheim2021}. 
A common feature of these works is that the optimal path equations are manifestly integrable and that the two-time boundary conditions present some symmetry property.
These characteristics are not shared by the MFT equations for SEP and this poses a major challenge.

However, the novel non-local transformation that we have discovered,  
\begin{align}
u(x,t) &= \frac{1}{\sigma'(\rho)}\frac{\partial}{\partial x}\sigma(\rho)\exp\left[-\int_{-\infty}^x dy\frac{\sigma'(\rho)}{2}\partial_yH\right],
\label{eq:u} \\
v(x,t) &= -\frac{2}{\sigma'(\rho)}\frac{\partial}{\partial x}\exp\left[\int_{-\infty}^x dy\frac{\sigma'(\rho)}{2}\partial_yH\right],
\label{eq:v}
\end{align}
allows us to map the MFT equations ~\eqref{eq:MFTrho} and \eqref{eq:MFTH}
to the AKNS equations
\cite{ablowitz1974}:
\begin{align}
\partial_tu(x,t) &= \partial_{xx}u(x,t)-2u(x,t)^2v(x,t), 
\label{eq:AKNSu}\\
\partial_tv(x,t) &= -\partial_{xx}v(x,t)+2u(x,t)v(x,t)^2.
\label{eq:AKNSv}
\end{align}
The transformation (\ref{eq:u}), (\ref{eq:v}) unveils the 
integrability of SEP at the hydrodynamic level (as was foreseen in \cite{Polychronakos2020} by finding 
solitons in the MFT equations).
In the low density limit, obtained by writing $\rho := \alpha \rho$ with $\alpha \to 0$, this change of variable reduces to the canonical Cole-Hopf transformation, i.e., $(u,v)\rightarrow(\partial_x\rho e^{-H},-\partial_xe^H$) and the AKNS equations decouple into two diffusion equations, evolving forward and backward in time, that were used to investigate reflecting Brownian motions \cite{Gerschenfeld2009,KMS2014}. The transformation above is valid for general quadratic $\sigma(\rho)$. The AKNS system with the same type of boundary conditions below also appeared in the analysis of \cite{LeDoussal2021a,Bettelheim2021}.

The  initial and final conditions of the MFT equations for SEP given in Eqs.~\eqref{eq:ICforH} and \eqref{eq:FCforH} are translated into the ones in terms of the AKNS variables by \eqref{eq:u} and \eqref{eq:v}. For the step initial density, they become 
\begin{align}
u(x,0) &= \omega\delta(x),
\label{eq:ICforu} \\
v(x,T) &= \delta(x).
\label{eq:FCforv}
\end{align}
Indeed, we obtain $u(x,t)\propto\partial_x\rho-\rho(1-\rho)\partial_xH$ and $v(x,t)\propto-\partial_xH$ showing that $u(x,0)$ and $v(x,T)$ are  proportional to the Dirac delta function.
Because the AKNS equations are invariant by the rescalings $u \to K u$ and $v \to K^{-1} v$, the amplitude of the Dirac delta function in Eq.~\eqref{eq:FCforv} can be set to unity by duly choosing the constant $K$. 
The parameter $\omega$ will be identified as the ubiquitous SEP parameter \cite{Gerschenfeld2009Bethe,IMScmp}:
\begin{equation}
\omega=(e^\lambda-1)\rho_-(1-\rho_+)+(e^{-\lambda}-1)\rho_+(1-\rho_-),  
\label{eq:omega}
\end{equation}
once the scattering amplitudes are determined in \eqref{eq:SA0-} and \eqref{eq:SAT+} 
(see \cite{SUPPL}). 
Using again that $\partial_xH(x,T)$ is a Dirac delta function at time $T$ from Eq.~\eqref{eq:FCforH}, and taking the rescaling factor $K$ into account, we observe that the transformations~\eqref{eq:u} and \eqref{eq:v} imply  
\begin{align}
u(x,T) &= 
\begin{cases}
\quad K\partial_x\rho(x,T), & \quad\quad\quad x<0,\\
\quad Ke^{-\Lambda}\partial_x\rho(x,T), & \quad\quad\quad x>0.
\end{cases} \label{eq:utorho}
\end{align}
The quantity $\Lambda = \frac12\int_{-\infty}^{+\infty} dy\sigma'(\rho)\partial_yH(y,T)$ is conserved by the MFT dynamics and is given by (see \cite{SUPPL})
\begin{equation}
e^{\Lambda}=e^{\lambda}\frac{1+(e^{-\lambda}-1)\rho_+}{1+(e^\lambda-1)\rho_-}.
\label{eq:LAMBDA}
\end{equation}

Similarly, at $t=0$, noting from  Eq.~\eqref{eq:ICforH} that $\partial_xH(x,0) = 2 \sigma(\rho)^{-1}\partial_x\rho(x,0)$ for $x \neq 0$
and using that $\Lambda$ is conserved, we deduce
\begin{align}
v(x,0) &= 
\begin{cases}
-2K^{-1}\sigma(\rho_-)^{-1}\partial_x\rho(x,0), & x<0,\\
-2K^{-1}\sigma(\rho_+)^{-1}e^{\Lambda}\partial_x\rho(x,0), & x>0.
\end{cases}
\label{eq:vtorho}
\end{align}
The value of $K$ will be determined after the AKNS equations have been solved.

The AKNS equations are classically integrable and the associated auxiliary linear problem \cite{Ablowitz1981} takes the form:
\begin{align}
\frac{\partial}{\partial x}\Psi(x,t) &= U(x,t;k)\Psi(x,t),
\label{eq:ZSU} \\
\frac{\partial}{\partial t}\Psi(x,t) &= V(x,t;k)\Psi(x,t),
\label{eq:ZSV}
\end{align}
where the vector $\Psi(x,t)$ plays the role of a   wave-function.
The $2\times2$ matrix-valued functions $U$ and $V$ are given by 
\begin{align}
U &=
\begin{pmatrix}
-ik & v(x,t) \\
u(x,t) & ik
\end{pmatrix},
\label{eq:U} \\
V&=
\begin{pmatrix}
2k^2+ u(x,t)v(x,t) & 2ikv(x,t)-\partial_xv(x,t) \\
2iku(x,t)+\partial_xu(x,t) & -2k^2- u(x,t)v(x,t)
\end{pmatrix}.
\label{eq:V}
\end{align}
The compatibility of Equations (\ref{eq:ZSU}) and  (\ref{eq:ZSV}) (i.e. $\partial_t\partial_x\Psi = \partial_x\partial_t\Psi$) is ensured by the zero curvature condition, $\frac{\partial U}{\partial t}-\frac{\partial V}{\partial x}+\left[U,V\right]=0$, which is met if the functions $u$ and $v$ satisfy  the AKNS system \eqref{eq:AKNSu} and \eqref{eq:AKNSv}.

Following the standard procedure of ISM \cite{Ablowitz1981,faddeev1987}, we first solve the direct scattering problem of the linear Equation \eqref{eq:ZSU}.
Assuming that $u(x,t)$ and $v(x,t)$ are rapidly decreasing functions, i.e. $u(x,t), v(x,t)\rightarrow0$ as $|x|\rightarrow\infty$, asymptotic states are  well-defined and the solutions behave as plain waves for large $|x|$.
The incoming/outgoing plane waves from $x\rightarrow-\infty$
\begin{align}
\phi(x;k)\sim
\begin{pmatrix}
e^{-ikx} \\
0
\end{pmatrix}
\quad {\rm and} \quad
\bar\phi(x;k)\sim
- \begin{pmatrix}
0 \\
e^{ikx}
\end{pmatrix},
\label{eq:BC-}
\end{align}
will be scattered at $x\rightarrow+\infty$ as follows
\begin{equation}
\phi(x;k) \sim
\begin{pmatrix}
a(k)e^{-ikx} \\
b(k)e^{ikx}
\end{pmatrix} 
\quad {\rm and} \quad
\bar\phi(x;k) \sim
\begin{pmatrix}
\bar{b}(k)e^{-ikx} \\
-\bar{a}(k)e^{ikx}
\end{pmatrix}. 
\label{eq:asyphi}
\end{equation}
This defines the scattering amplitudes, denoted by $a(k), \bar{a}(k), b(k), \bar{b}(k)$.
The calculation of these amplitudes at $t=0$ and $t=T$ under the initial and final conditions \eqref{eq:ICforu} and \eqref{eq:FCforv} is elementary, akin to solving the Schr\"odinger equation with a delta potential (see \cite{SUPPL}).
At $t=0$, we obtain 
\begin{align}
\label{eq:SA0-}
a(k,0) &= 1+\omega\hat{v}_+(k), \quad b(k,0)=\omega,
\\ \notag %\label{eq:SA0+} 
\bar a(k,0) &= 1+\omega\hat{v}_-(k), \quad \bar b(k,0)=-\left[\hat{v}(k)+\omega\hat{v}_+(k)\hat{v}_-(k)\right].
\end{align}
Similarly, at $t=T$, we have
\begin{align}
a(k,T) &= 1+\hat{u}_+(k), \quad b(k,T)=\hat{u}(k)+\hat{u}_+(k)\hat{u}_-(k), \notag
 \\ \label{eq:SAT+}
\bar a(k,T) &= 1+\hat{u}_-(k), \quad \bar b(k,T)=-1.
\end{align}
Here $\hat{u}_\pm(k)$ and $\hat{v}_\pm(k)$ are the half-Fourier transforms of $u(x,T)$ and $v(x,0)$ defined as
\begin{align}
\hat{u}_\pm(k) &= \int_\mathbb{R_\mp}u(x,T)e^{-2ikx}dx,
\label{eq:uhat} \\
\hat{v}_\pm(k) &= \int_\mathbb{R_\pm} v(x,0)e^{2ikx}dx
\label{eq:vhat},
\end{align}
and we use the notations $\hat{u}(k):=\hat{u}_+(k)+\hat{u}_-(k)$ and $\hat{v}(k):=\hat{v}_+(k)+\hat{v}_-(k)$. 

On the other hand, by combining Eq.~\eqref{eq:asyphi} with Eq.~\eqref{eq:ZSV}, the time evolution of the scattering amplitudes is obtained explicitly \cite{Ablowitz1981}
\begin{align}
a(k,t) &= a(k,0), \quad b(k,t)=b(k,0)e^{-4k^2t}, \label{evol:ab} \\
\bar a(k,t) &= \bar a(k,0), \quad \bar b(k,t)=\bar b(k,0)e^{4k^2t}. 
\label{evol:abBAR} 
\end{align}
The fact that the dynamics drastically simplifies in terms of the scattering amplitudes is a key feature of the ISM.
By using the evolution of $b(k,t)$ from $t=0$ to $T$, we deduce a closed equation for $\hat{u}_\pm(k)$: 
\begin{equation}
\hat{u}(k)+\hat{u}_+(k)\hat{u}_-(k)=\omega e^{-4k^2T}.
\label{eq:Eqforb}
\end{equation}
This equation is the Fourier transform of the equation for determining the density profile at final time, conjectured by Grabsch et al.  \cite{Benichou21c}, by ingenious microscopic considerations and inspection. Here in this paper we have shown that it arises as a simple consequence of the AKNS equations.
Note that a closely related relation also appears in the analysis of the KMP model by Bettelheim  et al. \cite{Bettelheim2021}. 
Rewriting Eq.~\eqref{eq:Eqforb} as
\begin{equation}
\left[\hat{u}_+(k)+1\right]\left[\hat{u}_-(k)+1\right]=1+\omega e^{-4k^2T},
\end{equation}
we obtain a scalar Riemann–Hilbert factorization problem of finding two functions, analytic on the upper (respectively lower) complex plane, with a given product along a specific contour.
The solution is standard \cite{Dunajski2009}, by taking the logarithm of Eq.~\eqref{eq:Eqforb} and using the Cauchy Transform with an infinitesimal constant $\epsilon>0$ 
\begin{equation}
\label{uhatsol}
\hat{u}_\pm(k)+1=\exp\left[\pm\frac{1}{2\pi i}\int_{-\infty}^\infty\frac{\log(1+\omega e^{-4q^2T})}{q-k\mp i\epsilon}dq\right].
\end{equation}
The coefficient in front of the exponential on the right hand side of \eqref{uhatsol} is taken to be unity to ensure that $\hat{u}_\pm(k)$ vanishes for $k\to\infty$ for bounded $u$.
Expanding the logarithm inside the integral and using the following formula (see e.g. Eqs. 7.2.3 and 7.7.2 in \cite{NIST:DLMF}) 
\begin{equation}
\pm\frac{1}{\pi i}\int_{-\infty}^\infty\frac{e^{-q^2}}{q-k\mp i\epsilon}dq=e^{-k^2}\mathrm{erfc}(\mp ik), 
\end{equation}
we conclude that \cite{Benichou21c}
\begin{equation}
\hat{u}_\pm(k)+1=\exp\left[-\frac{1}{2}\sum_{n=1}^\infty\frac{(-\omega e^{-4k^2T})^n}{n}\mathrm{erfc}(\mp i\sqrt{4nT}k)\right]
\label{eq:soluhat}
\end{equation}
where $\mathrm{erfc}(x)$ is the complementary error function.
Since $a(k,t)$ and $\bar a(k,t)$ are conserved, see Eqs~\eqref{evol:ab},~\eqref{evol:abBAR}, we find from  Eqs~\eqref{eq:SA0-}--\eqref{eq:SAT+} that
$\omega\hat{v}_\pm = \hat{u}_\pm$.

Therefore, the density profiles at $t=0$ and $T$ are  determined up to the constant $K$ by integrating Eqs.~\eqref{eq:utorho} and \eqref{eq:vtorho} with  $\rho(x,t)\rightarrow\rho_\pm$ for $x\rightarrow\pm\infty$.
Finally, $K$ is fixed by imposing the total
mass conservation $\int_{-\infty}^\infty [\rho(x,T)-\rho(x,0)]dx=0$ (See \cite{SUPPL}):
\begin{equation}
  K = -2\sinh (\lambda/2) e^{\Lambda/2},
  \label{eq:K}
\end{equation}
where $\Lambda$ is given in Eq.~\eqref{eq:LAMBDA}.

\begin{figure}[ht]
\begin{minipage}[b]{0.45\linewidth}
   \hspace{-0.8cm}
    \includegraphics[scale=.47]{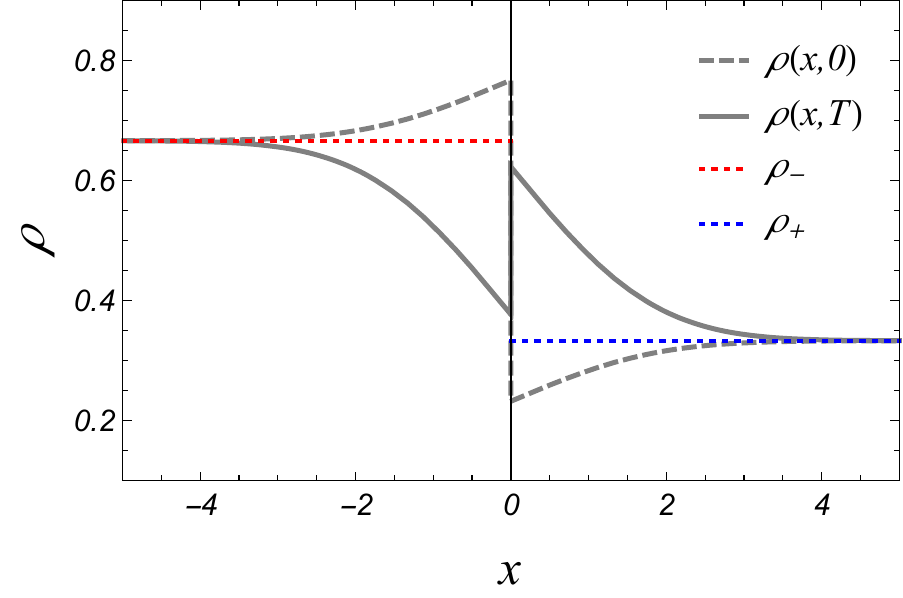}
\end{minipage}
  \begin{minipage}[b]{0.45\linewidth}
  %\hspace*{-0.3cm}
  \includegraphics[scale=.47]{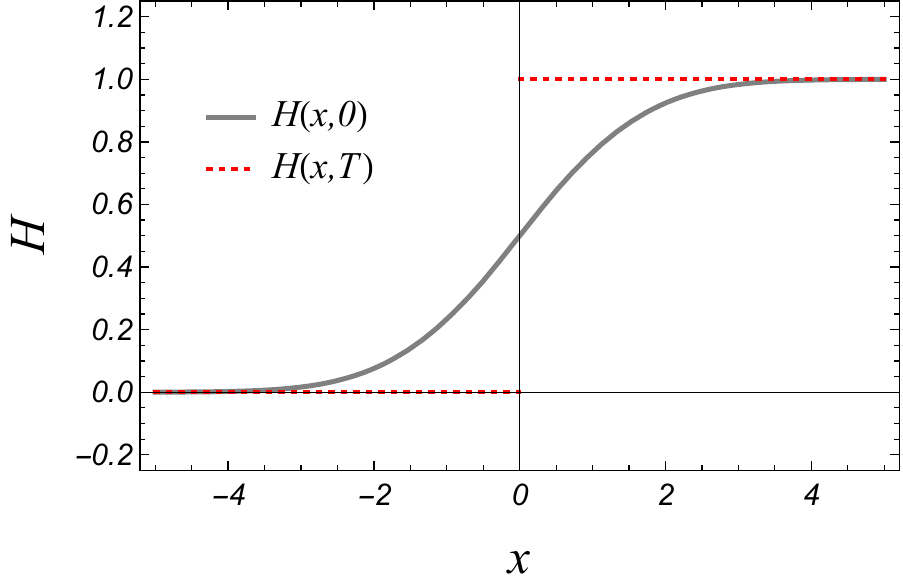}
\end{minipage}
    \caption{Optimal profiles of $\rho$ (left) and $H$ (right) at $t=0$ and at $t=T$, with  $\rho_+=1/3$, $\rho_-=2/3$, $\lambda=1$ and $T=1$.}
    \label{fig:rhoH}
\end{figure}

Combining all the calculations above, we present exact formulas for the density profiles.
The optimal density fluctuation at initial time, $t = 0$, is given by
\begin{equation}
\rho(x,0) =
\begin{cases}
\rho_-+A_-\int_{-\infty}^xv(y,0)dy, \quad x<0, \\
\rho_++A_+\int_x^\infty v(y,0)dy, \quad x>0,
\end{cases}
\label{eq:rho0} 
\end{equation}
with $ A_{\pm}= \sigma(\rho_{\pm}) 
\frac{e^{\mp\lambda}-1}{2}
\sqrt{\frac{1+(e^{\pm\lambda}-1)\rho_\mp}{1+(e^{\mp\lambda}-1)\rho_\pm}} .$
% \begin{align}
% \rho(x,0) &= \rho_-+A_-\int_{-\infty}^xv(y,0)dy, \quad x<0, \label{eq:rho0-} \\
% \rho(x,0) &= \rho_++A_+\int_x^\infty v(y,0)dy, \quad x>0,\label{eq:rho0+} 
% \end{align}
% with $ A_{\pm}= \sigma(\rho_{\pm}) 
% \frac{e^{\mp\lambda}-1}{2}
At $t=T$, the form of the profile is  obtained by replacing   $v(y,0)$ by $u(y,T)$ and $A_{\pm}$ by $B_{\pm}= - \frac{\sigma(\rho_{\pm})}{2 A_{\pm}}$ in
Eq.~\eqref{eq:rho0}.
%Eqs.~\eqref{eq:rho0-} and \eqref{eq:rho0+}.
The response field $H$ at $t=0$ is determined thanks to \eqref{eq:ICforH}.
While the final density can be extracted from the information in \cite{Benichou21c}, our scheme using ISM allows us to determine simultaneously the optimal profiles of $\rho$ and $H$ exactly at both initial and final time.
%The initial and final profiles of both the fields $\rho$ and $H$ 
An example for all of them is represented in Fig. \ref{fig:rhoH}. 

Finally, the cumulant generating function is retrieved by noting that $\mu$ and $Q_T$ are dual by Legendre transform i.e. $\frac{d\mu}{d\lambda} = Q_T/\sqrt{T}$
\cite{Bettelheim2021,Dandekar}.
Calculating the total current $Q_T$ from the profiles at $t=0$ and $t=T$, given in Eq.~\eqref{eq:rho0} and beneath, we obtain, using $\mu(0)=0$ and \eqref{eq:omega},
\begin{equation}
\mu(\lambda)=\frac{1}{\sqrt{\pi}}\sum_{n=1}^\infty\frac{(-1)^{n-1}\omega^n}{n^{3/2}}. 
\label{eq:SCGF}
\end{equation}
This formula was first found in  \cite{Gerschenfeld2009Bethe}
at the microscopic level by applying the Bethe ansatz to the SEP.
Here it has been deduced from the action principle of the MFT.
The large deviations of a tracer particle, derived  microscopically in \cite{IMSprl,IMScmp}, can be extracted along similar lines from the MFT framework \cite{MMSprep}.
 
To summarize, we have presented the first exact solution of the time-dependent MFT equations for SEP.
Albeit these equations were known for a long time, their  solution had remained out of reach due to their intrinsic complexity and to cumbrous boundary conditions with respect to time. 
A key to the solution has been our novel nonlocal change of variables given in Eqs.~\eqref{eq:u} and \eqref{eq:v}, that generalizes the canonical Cole-Hopf transformation.
This enabled us to map the MFT equations to the integrable AKNS system and to use the inverse scattering method.
We have derived exact expressions for the optimal density profile and the response field both at initial and final times.
By retrieving the cumulant generating function of the integrated current, previously found by a microscopic calculation, we have provided a first analytic confirmation of the validity of the macroscopic fluctuation theory in time dependent regime.

The present work can be extended in multiple directions.
Many variants of the exclusion process -- different    geometries, initial conditions, multiple species, asymmetry, tagged particles, defects...-- have been explored during the last decades and ought to be analyzed with the MFT.
Besides, some diffusive interacting particle processes out of equilibrium, with identical transport coefficients,   could be solvable by ISM at the macroscopic level, though the corresponding microscopic models may not be integrable.
Understanding the connections between different scales of description and various forms of integrability poses challenging problems.

We are convinced that the analysis of the Macroscopic Fluctuation Theory with Inverse Scattering, applied to the KMP model \cite{Bettelheim2021}, to SEP in the present work and to the closely related KPZ equation subject to weak noise \cite{LeDoussal2021a,LeDoussal2021b}, opens a  fascinating new perspective in the study of dynamical  fluctuations in systems far of equilibrium. 

\vspace{2mm}

\noindent{\bf Acknowledgments}
K.M. is thankful to S. Mallick for a careful reading of the manuscript and to A. Grabsch, O. B\'enichou, R. Dandekar and P. L. Krapivsky for discussions. The work of K.M. has been supported by the project RETENU ANR-20-CE40-0005-01 of the French National Research Agency (ANR).
The work of T.S. has been supported by JSPS KAKENHI Grants No. JP16H06338, No. JP18H01141, No. JP18H03672, No. JP19L03665, No. JP21H04432.

\bibliographystyle{unsrt}
\bibliography{MMS.bib}

\begin{thebibliography}{10}

\bibitem{OnsMach1}
L.~Onsager and S.~Machlup.
\newblock {Fluctuations and Irreversible Processes}.
\newblock {\em Phys. Rev.}, 91:1505--1512, 1953.

\bibitem{OnsMach2}
S.~Machlup and L.~Onsager.
\newblock {Fluctuations and Irreversible Process. II. Systems with Kinetic
  Energy}.
\newblock {\em Phys. Rev.}, 91:1512--1515, 1953.

\bibitem{Varadhan1984}
S.~R.~S. Varadhan.
\newblock {\em Large Deviations and Applications}.
\newblock Siam, Philadelphia, 1984.

\bibitem{TOUCHETTE}
H.~Touchette.
\newblock {The Large Deviation Approach to Statistical Mechanics}.
\newblock {\em Phys. Rep.}, 478:1--69, 2009.

\bibitem{LebowitzSpohn1999}
J.~L. Lebowitz and H.~Spohn.
\newblock {A Gallavotti-Cohen-Type Symmetry in the Large Deviation Functional
  for Stochastic Dynamics}.
\newblock {\em J. Stat. Phys.}, 95:333--365, 1999.

\bibitem{Derrida2007}
B.~Derrida.
\newblock Non-equilibrium steady states: fluctuations and large deviations of
  the density and of the current.
\newblock {\em J. Stat. Mech.}, P07023, 2007.

\bibitem{Derrida2011}
B.~Derrida.
\newblock Microscopic versus macroscopic approaches to non-equilibrium systems.
\newblock {\em J. Stat. Mech.}, P01030, 2011.

\bibitem{TouchetteHARRIS}
H.~Touchette and R.~J. Harris.
\newblock {\em Large Deviation Approach to Nonequilibrium Systems in
  Nonequilibrium Statistical Physics of Small Systems: Fluctuation Relations
  and Beyond (R. Klages, W. Just and C. Jarzynski Eds.)}.
\newblock Wiley‐VCH Verlag, 2013.

\bibitem{GianniBrazil}
L.~Bertini, A.~De~Sole, D.~Gabrielli, G.~Jona-Lasinio, and C.~Landim.
\newblock Large deviation approach to non equilibrium processes in stochastic
  lattice gases.
\newblock {\em Bull Braz Math Soc.}, 37(4):611--643, 2006.

\bibitem{Bertini2001}
L.~Bertini, A.~De~Sole, D.~Gabrielli, G.~Jona-Lasinio, and C.~Landim.
\newblock {Fluctuations in Stationary Nonequilibrium States of Irreversible
  Processes}.
\newblock {\em Phys. Rev. Lett.}, 87:040601, 2001.

\bibitem{Bertini2002}
L~Bertini, A~De~Sole, D~Gabrielli, G.~Jona-Lasinio, and C~Landim.
\newblock {Macroscopic Fluctuation Theory for Stationary Non-Equilibrium
  States}.
\newblock {\em J. Stat. Phys.}, 107:635--675, 2002.

\bibitem{Bodineau2004}
T.~Bodineau and B.~Derrida.
\newblock {Current Fluctuations in Nonequilibrium Diffusive Systems: An
  Additivity Principle}.
\newblock {\em Phys. Rev. Lett.}, 92:180601, 2004.

\bibitem{Bodineau05}
T.~Bodineau and B.~Derrida.
\newblock Distribution of current in non-equilibrium diffusive systems and
  phase transitions.
\newblock {\em Phys. Rev. E}, 72:066110, 2005.

\bibitem{Bertini2015}
L.~Bertini, A.~De~Sole, D.~Gabrielli, G.~Jona-Lasinio, and C.~Landim.
\newblock Macroscopic fluctuation theory.
\newblock {\em Rev. Mod. Phys.}, 87:593--636, 2015.

\bibitem{Jona-Lasinio2014}
G.~Jona-Lasinio.
\newblock Thermodynamics of stationary states.
\newblock {\em J. Stat. Mech.}, P02004, 2014.

\bibitem{KLS1984}
S.~Katz, J.~L. Lebowitz, and H.~Spohn.
\newblock Nonequilibrium steady states of stochastic lattice gas models of fast
  ionic conductors.
\newblock {\em J. Stat. Phys.}, 34:497--537, 1984.

\bibitem{Spohn1991}
H.~Spohn.
\newblock {\em Large Scale Dynamics of Interacting Particles}.
\newblock Springer-Verlag, New York, 1991.

\bibitem{SchutzReview}
G.~M. Schütz.
\newblock Exactly solvable models for many-body systems far from equilibrium.
\newblock {\em Phase Transitions and Critical Phenomena}, 19:1--251, 12 2001.

\bibitem{DerridaPhysREP}
B.~Derrida.
\newblock {An exactly soluble non-equilibrium system: The asymmetric simple
  exclusion process}.
\newblock {\em {Physics Reports}}, 301(1):65--83, 1998.

\bibitem{CHOU2011}
T.~Chou, K.~Mallick, and R.~K.~P. Zia.
\newblock Non-equilibrium statistical mechanics: from a paradigmatic model to
  biological transport.
\newblock {\em Reports on Progress in Physics}, 74(11):116601, Oct 2011.

\bibitem{Derrida-Leb01}
B.~Derrida, J.~L. Lebowitz, and E.~R. Speer.
\newblock {Free Energy Functional for Nonequilibrium Systems: An Exactly
  Solvable Case}.
\newblock {\em Phys. Rev. Lett.}, 87:150601, 2001.

\bibitem{Derrida-Leb02a}
B.~Derrida, J.~L. Lebowitz, and E.~R. Speer.
\newblock {Exact Free Energy Functional for a Driven Diffusive Open Stationary
  Nonequilibrium System}.
\newblock {\em Phys. Rev. Lett.}, 89:030601, 2002.

\bibitem{Derrida-Leb02b}
B.~Derrida, J.~L. Lebowitz, and E.~R. Speer.
\newblock {Large Deviation of the Density Profile in the Steady State of the
  Open Symmetric Simple Exclusion Process}.
\newblock {\em J. Stat. Phys.}, 107:599--634, 2002.

\bibitem{Bertini2003SEP}
L.~Bertini, A.~De~Sole, D.~Gabrielli, G.~Jona-Lasinio, and C.~Landim.
\newblock {Large Deviations for the Boundary Driven Symmetric Simple Exclusion
  Process}.
\newblock {\em Math. Physics, Anal. Geom}, 6(3):231--267, 2003.

\bibitem{Gerschenfeld2009Bethe}
B.~Derrida and A.~Gerschenfeld.
\newblock {Current Fluctuations of the One Dimensional Symmetric Simple
  Exclusion Process with Step Initial Condition}.
\newblock {\em J. Stat. Phys.}, 136:1--15, 2009.

\bibitem{Gerschenfeld2009}
B.~Derrida and A.~Gerschenfeld.
\newblock {Current Fluctuations in One Dimensional Diffusive Systems with a
  Step Initial Density Profile}.
\newblock {\em J. Stat. Phys.}, 137:978--1000, 2009.

\bibitem{IMSprl}
T.~Imamura, K.~Mallick, and T.~Sasamoto.
\newblock {Large Deviations of a Tracer in the Symmetric Exclusion Process}.
\newblock {\em Phys. Rev. Lett.}, 118:160601, 2017.

\bibitem{IMScmp}
T.~Imamura, K.~Mallick, and T.~Sasamoto.
\newblock {Distribution of a Tagged Particle Position in the One-Dimensional
  Symmetric Simple Exclusion Process with Two-Sided Bernoulli Initial
  Condition}.
\newblock {\em Commun. Math. Phys.}, 384:1409, 2021.

\bibitem{Kipnis89}
C.~Kipnis, S.~Olla, and S.~R.~S. Varadhan.
\newblock Hydrodynamics and large deviations for simple exclusion processes.
\newblock {\em Comm. Pure Appl. Math.}, 42:115--137, 1989.

\bibitem{SethuramanVaradhan2013}
{S. Sethuraman, S.R.S. Varadhan}.
\newblock Large deviations for the current and tagged particle in 1d symmetric
  simple exclusion.
\newblock {\em Ann. Prob.}, 41:1461--1512, 2013.

\bibitem{Schutz1997}
G.~M. Schütz.
\newblock Exact solution of the master equation for the asymmetric exclusion
  process.
\newblock {\em Journal of Statistical Physics}, 88(1-2):427–445, 1997.

\bibitem{TW2008a}
C.~A. Tracy and H.~Widom.
\newblock {Integral Formulas for the Asymmetric Simple Exclusion Process}.
\newblock {\em Com. Math. Phys.}, 279:815--844, 2008.

\bibitem{BCS2014}
{A. Borodin, I. Corwin, T. Sasamoto}.
\newblock {From duality to determinants for $q$-TASEP and ASEP}.
\newblock {\em Ann. Prob.}, 42:2314--2382, 2014.

\bibitem{BC2014}
A.~Borodin and I.~Corwin.
\newblock {Macdonald processes}.
\newblock {\em Prob. Th. Rel. Fields}, 158:225--400, 2014.

\bibitem{Bodineau07}
T.~Bodineau and B.~Derrida.
\newblock Cumulants and large deviations of the current through non-equilibrium
  steady states.
\newblock {\em C. R. Physique}, 8:540--555, 2007.

\bibitem{Krapivsky2012}
P.~L. Krapivsky and Baruch Meerson.
\newblock Fluctuations of current in nonstationary diffusive lattice gases.
\newblock {\em Phys. Rev. E}, 86:031106, 2012.

\bibitem{KMS2014}
P.~L. Krapivsky, K.~Mallick, and T.~Sadhu.
\newblock Large deviations in single-file diffusion.
\newblock {\em Phys. Rev. Lett.}, 113:078101, 2014.

\bibitem{Meerson13}
B.~Meerson and P.~V. Sasorov.
\newblock Extreme current fluctuations in a non-stationary stochastic heat
  flow.
\newblock {\em J. Stat. Mech.}, 12:P12011, 2013.

\bibitem{Sasorov2014}
B.~Meerson and P.~V. Sasorov.
\newblock Extreme current fluctuations in lattice gases: Beyond nonequilibrium
  steady states.
\newblock {\em Phys. Rev. E}, 89:010101, 2014.

\bibitem{Kamenev16b}
M.~Janas, A.~Kamenev, and B.~Meerson.
\newblock {Dynamical phase transition in large-deviation statistics of the
  Kardar-Parisi-Zhang equation}.
\newblock {\em Phys. Rev. E}, 94:032133, 2016.

\bibitem{LeDoussal2021a}
Krajenbrink A. and P.~Le~Doussal.
\newblock {The inverse scattering of the Zakharov-Shabat system solves the weak
  noise theory of the Kardar-Parisi-Zhang equation}.
\newblock {\em Phys. Rev. Lett}, 127:064101, 2021.

\bibitem{LeDoussal2021b}
A.~Krajenbrink and P.~Le Doussal.
\newblock {Inverse scattering solution of the weak noise theory of the
  Kardar-Parisi-Zhang equation with flat and Brownian initial conditions}.
\newblock {\em arXiv:2107.13497}, 2021.

\bibitem{Bettelheim2021}
E.~Bettelheim, N.~R. Smith, and B.~Meerson.
\newblock {Inverse Scattering Method Solves the Problem of Full Statistics of
  Nonstationary Heat Transfer in the Kipnis-Marchioro-Presutti Model}.
\newblock {\em arXiv:2112.02474}, 2021.

\bibitem{Benichou21c}
A.~Grabsch, A.~Poncet, P.~Rizkallah, P.~Illien, and O.~B\'{e}nichou.
\newblock {Exact Closure and Solution for Spatial Correlations in Single-File
  Diffusion}.
\newblock {\em To appear in Science Adv., arXiv:2110.09269}, 2021.

\bibitem{Illien2013}
P.~Illien, O.~B\'{e}nichou, C.~Mej\'{\i}a-Monasterio, G.~Oshanin, and
  R.~Voituriez.
\newblock {Active Transport in Dense Diffusive Single-File Systems}.
\newblock {\em Phys. Rev. Lett.}, 111:038102, 2013.

\bibitem{Benichou21a}
A.~Poncet, O.~B\'{e}nichou, and P.~Illien.
\newblock Cumulant generating functions of a tracer in quenched dense symmetric
  exclusion processes.
\newblock {\em Phys. Rev. E}, 103:L040103, 2021.

\bibitem{Benichou21b}
A.~Poncet, A.~Grabsch, P.~Illien, and O.~B\'enichou.
\newblock {Generalized Correlation Profiles in Single-File Systems}.
\newblock {\em Phys. Rev. Lett.}, 127:220601, 2021.

\bibitem{krapivskyBook}
P.~L. Krapivsky, S.~Redner, and E.~Ben-Naim.
\newblock {\em A Kinetic View of Statistical Physics}.
\newblock Cambridge University Press, 2010.

\bibitem{KMS_tagged}
P.~L. Krapivsky, K.~Mallick, and T.~Sadhu.
\newblock {Tagged Particle in Single-File Diffusion}.
\newblock {\em J. Stat. Phys.}, 160:885--925, 2015.

\bibitem{Vilenkin14}
A.~Vilenkin, B.~Meerson, and P.~V. Sasorov.
\newblock Extreme fluctuations of current in the symmetric simple exclusion
  process: a non-stationary setting.
\newblock {\em J. Stat. Mech.}, 06:P06007, 2014.

\bibitem{Ablowitz1981}
M.~J. Ablowitz and H.~Segur.
\newblock {\em Solitons and the Inverse Scattering Transform}.
\newblock SIAM, Philadelphia, 1981.

\bibitem{faddeev1987}
L.~Faddeev and L.~Takhtajan.
\newblock {\em {Hamiltonian Methods in the Theory of Solitons}}.
\newblock Springer Berlin Heidelberg, 1987.

\bibitem{ablowitz1974}
M.~J. Ablowitz, D.~J Kaup, A.~C Newell, and H.~Segur.
\newblock The inverse scattering transform-fourier analysis for nonlinear
  problems.
\newblock {\em Stud. Appl. Math.}, 53(4):249--315, 1974.

\bibitem{Polychronakos2020}
A.~P. Polychronakos.
\newblock Solitons in fluctuating hydrodynamics of diffusive processes.
\newblock {\em Phys. Rev. E}, 101:022209, 2020.

\bibitem{SUPPL}
Supplementary material.

\bibitem{Dunajski2009}
M.~Dunajski.
\newblock {\em Solitons, Instantons and Twistors}.
\newblock OUP, Oxford, 2009.

\bibitem{NIST:DLMF}
{\it NIST Digital Library of Mathematical Functions}.
\newblock http://dlmf.nist.gov/, Release 1.1.4 of 2022-01-15.
\newblock F.~W.~J. Olver, A.~B. {Olde Daalhuis}, D.~W. Lozier, B.~I. Schneider,
  R.~F. Boisvert, C.~W. Clark, B.~R. Miller, B.~V. Saunders, H.~S. Cohl, and
  M.~A. McClain, eds.

\bibitem{Dandekar}
R.~Dandekar.
\newblock Private communication.

\bibitem{MMSprep}
K.~Mallick, H.~Moriya, and T.~Sasamoto.
\newblock (in preparation).
\newblock 2022.

\end{thebibliography}

\newpage

\begin{center}
{\bf Supplementary Material}
\end{center}

\section{A. Scattering amplitudes}
\label{sec:AppendixA}
\setcounter{equation}{0}
\renewcommand{\theequation}{A.\arabic{equation}}
Here, we explain briefly how to get the scattering amplitudes given in Eqs.~\eqref{eq:SA0-} and \eqref{eq:SAT+}. The starting point is Eq.~\eqref{eq:ZSU} as usual. For convenience, let us introduce the components of the wave-functions such as
\begin{equation}
\phi(x;k)=
\begin{pmatrix}
\phi_1(x) \\
\phi_2(x)
\end{pmatrix},
\quad
\bar\phi(x;k)=
\begin{pmatrix}
\bar\phi_1(x) \\
\bar\phi_2(x)
\end{pmatrix}
\end{equation}
and remove the trivial oscillation (i.e.  we go to the interaction picture). Then, Eq.~\eqref{eq:ZSU} becomes
\begin{align}
& \quad \frac{\partial}{\partial x}
\begin{pmatrix}
e^{ikx}\phi_1(x) \\
e^{-ikx}\phi_2(x)
\end{pmatrix} \notag
 \\
 &=
 \begin{pmatrix}
0 & v(x,t)e^{2ikx} \\
u(x,t)e^{-2ikx} & 0
\end{pmatrix}
\begin{pmatrix}
e^{ikx}\phi_1(x) \\
e^{-ikx}\phi_2(x)
\end{pmatrix}.
\end{align}
Since the discussion is almost the same, we restrict ourselves to the  $t=0$
 case for  which the initial condition is given by Eq.~\eqref{eq:ICforu}. 
First, we solve
\begin{equation}
\partial_x\left[e^{-ikx}\phi_2(x)\right]=\omega\delta(x)e^{-ikx}\phi_1(x)
\end{equation}
under the boundary condition designated in Eq.~\eqref{eq:BC-}.
Plugging the solution $e^{-ikx}\phi_2(x)=\omega\phi_1(0)\theta(x)$ into the rest of the equation, we have
\begin{equation}
\partial_x\left[e^{ikx}\phi_1(x)\right]=\omega \phi_1(0)\theta(x)v(x,0)e^{2ikx}.
\end{equation}
Integrating this ODE under the boundary condition in Eq.~\eqref{eq:BC-} again, we obtain
\begin{equation}
e^{ikx}\phi_1(x)=1+\omega\phi_1(0)\theta(x)\int_0^x v(y,0)e^{2iky}dy.
\end{equation}
Setting $x=0$, we find that $\phi_1(0)=1$. Bringing it back to the expression of $\phi_1(x)$ and considering the limit $x\rightarrow+\infty$, the scattering amplitude for $a(k,0)$ can be extracted from the definition \eqref{eq:asyphi}. For $b(k,0)$, recalling the solution $e^{-ikx}\phi_2(x)=\omega\phi_1(0)\theta(x)$ and considering the limit $x\rightarrow+\infty$, we get the answer by definition \eqref{eq:asyphi} again. Following this procedure, we see Eq.~\eqref{eq:SA0-} is valid. Regarding the scattering amplitudes such as $\bar a(k,0)$ and $\bar b(k,0)$, they are extracted by solving the set of ODEs for $\bar\phi_1(x)$ and $\bar\phi_2(x)$ in the same fashion. The $t=T$ case  with the final condition given by Eq.~\eqref{eq:FCforv} is entirely similar.

\section{B. Determination of $\omega$ and $\Lambda$}
\label{sec:AppendixB}
\setcounter{equation}{0}
\renewcommand{\theequation}{B.\arabic{equation}}

Let us consider a $2\times2$ matrix-valued function $\Omega(x,t)$, which satisfies the $k=0$ case of Eq.~\eqref{eq:ZSU}, i.e.
\begin{equation}
\label{eq:OmegaU}
\frac{\partial}{\partial x}\Omega(x,t)=U(x,t;0)\Omega(x,t).
\end{equation}
$\Omega(x,t)$ is known to connect the AKNS system to the  Landau-Lifshitz equation \cite{faddeev1987}, though we do not use this fact in this paper. 
When $\Psi(x;t)$ is the solution to Eq.~\eqref{eq:ZSU} with $k=0$,  we have 
$\Psi(x;t)=\Omega(x,t)\Omega^{-1}(y,t) \Psi(y;t)$, because the right hand side satisfies the same equation and has the same value at $x=y$. In other words, 
$\Omega(x,t)\Omega^{-1}(y,t)$ connects the solutions at $x$ and $y$. If in particular we consider $\phi(x;k),\bar\phi(x;k)$ with $k=0$ and use the asymptotics in \eqref{eq:BC-} and \eqref{eq:asyphi}, 
we find the relation 
\begin{equation}
\lim_{\substack{x\to+\infty\\y\to-\infty}}\Omega(x,t)\Omega^{-1}(y,t)
=
\begin{pmatrix}
a(0,t) & -\bar{b}(0,t) \\
b(0,t) & ~~\bar{a}(0,t)
\end{pmatrix}.
\label{eq:Omega_ab}
\end{equation}

On the other hand, expressing $U(x,t;k)$ with $k=0$ in (\ref{eq:U}) in terms of $\rho,H$ using (\ref{eq:u}) and \eqref{eq:v}, the solution to  
the linear ODE (\ref{eq:OmegaU}) is  
given by
\begin{equation}
\Omega(x,t)=
\begin{pmatrix}
Ce^{\int_{-\infty}^x dy(1-\rho)\partial_yH} & Ce^{-\int_{-\infty}^x dy\rho\partial_yH} \\
-(1-\rho)e^{\int_{-\infty}^x dy\rho\partial_yH} & \rho e^{-\int_{-\infty}^x dy(1-\rho)\partial_yH}
\end{pmatrix},
\label{eq:Omega}
\end{equation}
with a constant $C$ independent of $x$. This can be checked by a direct substitution into \eqref{eq:OmegaU}.

Using this expression, together with the boundary conditions in terms of MFT variables,
\begin{align}
\rho(x,t)\sim\rho_-, &\quad H(x,t)\sim 0, 
\quad\mathrm{as}\quad x\rightarrow-\infty, \\
\rho(x,t)\sim\rho_+, &\quad H(x,t)\sim \lambda, \quad\mathrm{as}\quad x\rightarrow+\infty,
\end{align}
the left hand side of \eqref{eq:Omega_ab} can be evaluated to be in the form,
\begin{equation}
\begin{pmatrix}
C_1[1+(e^\lambda-1)\rho_-] & -C_2(e^\lambda-1) \\
-C_2^{-1}[r_--e^{-\lambda}r_+] 
&C_1^{-1}[1+(e^{-\lambda}-1)\rho_+]
\end{pmatrix}
\end{equation}
with $r_\pm = \rho_\pm(1-\rho_{\mp})$ and two constants given by  $C_1=e^{\Lambda/2-\lambda/2}$ and 
$C_2=CC_1$ where $\Lambda = \frac12\int_{-\infty}^{+\infty} dy\sigma'(\rho)\partial_yH(y,T)$ is the conserved quantity introduced below \eqref{eq:utorho}.  By taking the ratio of the diagonal elements, the quantity $\Lambda$ can readily be evaluated as
\begin{equation}
e^{\Lambda-\lambda}=\frac{1+(e^{-\lambda}-1)\rho_+}{1+(e^\lambda-1)\rho_-}.
\label{eq:Q}
\end{equation}

Multiplying the off-diagonal elements and using $b(0,0)=\omega$ in Eq.~\eqref{eq:SA0-} and $\bar{b}(0,0)=\bar{b}(0,T)=-1$ by Eqs.~\eqref{eq:SAT+} and \eqref{evol:abBAR}, one concludes that 
\begin{equation}
\omega=(e^\lambda-1)\rho_-(1-\rho_+)+(e^{-\lambda}-1)\rho_+(1-\rho_-).
\end{equation}

\section{C. Determination of $K$} 
\label{sec:AppendixC}
\setcounter{equation}{0}
\renewcommand{\theequation}{C.\arabic{equation}}

As mentioned in the main text, the constant $K$ is fixed by the mass conservation law
\begin{equation}
\int_{-\infty}^\infty[\rho(x,T)-\rho(x,0)]dx=0.
\end{equation}
We insert the integrated forms of Eqs.~\eqref{eq:utorho} and \eqref{eq:vtorho} into the left hand side and  integrate by parts.  
The resulting integrals proportional to $\int_\mathbb{R_\mp}xu(x,T)dx $ and $\int_\mathbb{R_\pm}xv(x,0)dx $ are identified to the derivatives of $\hat{u}(k)$ and $\hat{v}(k)$ at 0, as follows:
\begin{equation}
\int_\mathbb{R_\mp}xu(x,T)dx=\frac{\hat{u}'_\pm(0)}{-2i}=\pm\frac{\sqrt{T}}{\sqrt{\pi}}\mathrm{Li}_{1/2}(-\omega)\sqrt{1+\omega}
\end{equation}
and
\begin{equation}
\int_\mathbb{R_\pm}xv(x,0)dx=\frac{\hat{v}'_\pm(0)}{2i}=\mp\frac{\sqrt{T}}{\sqrt{\pi}}\mathrm{Li}_{1/2}(-\omega)\frac{\sqrt{1+\omega}}{\omega}
\end{equation}
where 
\begin{equation}
\mathrm{Li}_s(z)=\sum_{n=1}^\infty\frac{z^n}{n^s}
\end{equation}
is the polylogarithm of order $s$.
Using these formulae, we obtain
\begin{equation}
K^2=\frac{2\omega(e^\Lambda-1)}{\sigma(\rho_-)-\sigma(\rho_+)e^{-\Lambda}}=4\sinh^2(\lambda/2)e^\Lambda.
\end{equation}
When the square root is taken, the minus sign should be chosen, by considering that a positive current must be generated when $\lambda>0$ and $\rho_->\rho_+$. Hence, Eq.~\eqref{eq:K} is proved.

\end{document}